\documentclass{osa-article}

\journal{oe}

\usepackage{graphicx}
\usepackage{xcolor} 
\usepackage{dcolumn}
\usepackage{bm}
\usepackage{textgreek}
\usepackage[version = 4]{mhchem}
\usepackage{hyperref}

\usepackage{siunitx} 
\usepackage{comment} 

\begin{document}

\title{Designing arbitrary one-dimensional potentials on an atom chip}

\author{Mohammadamin Tajik,\authormark{1} Bernhard Rauer,\authormark{1,2} Thomas Schweigler,\authormark{1} Federica Cataldini,\authormark{1} Jo{\~a}o Sabino,\authormark{1,3,4} Frederik S. M{\o}ller,\authormark{1} Si-Cong Ji,\authormark{1} Igor E. Mazets,\authormark{1,5} and J{\"o}rg Schmiedmayer\authormark{1,*}}

\address{\authormark{1}Vienna Center for Quantum Science and Technology, Atominstitut, TU Wien, Stadionallee 2, 1020 Vienna, Austria\\
\authormark{2}Laboratoire Kastler Brossel, Ecole Normale Sup\'{e}rieure, Coll\`{e}ge de France, CNRS UMR 8552, Sorbonne universit\'{e}, 24 rue Lhomond, 75005 Paris, France \\
\authormark{3}Instituto de Telecomunica\c{c}\~{o}es, Physics of Information and Quantum Technologies Group, Av. Rovisco Pais 1, 1049-001, Lisbon, Portugal\\
\authormark{4}Instituto Superior T\'{e}cnico, Universidade de Lisboa, Av. Rovisco Pais 1, 1049-001, Lisbon, Portugal\\
\authormark{5}Wolfgang Pauli Institute c/o Fakult{\"a}t f{\"u}r Mathematik, Universit{\"a}t Wien, Oskar-Morgenstern-Platz 1, 1090 Vienna, Austria
}

\email{\authormark{*}schmiedmayer@atomchip.org} 



\begin{abstract}
We use laser light shaped by a digital micro-mirror device to realize arbitrary optical dipole potentials for one-dimensional (1D) degenerate Bose gases of \ce{^{87}_{}Rb} trapped on an atom chip. Superposing optical and magnetic potentials combines the high flexibility of optical dipole traps with the advantages of magnetic trapping, such as effective evaporative cooling and the application of radio-frequency dressed state potentials. As applications, we present a $160\, \si{\micro \meter}$ long box-like potential with a central tuneable barrier, a box-like potential with a sinusoidally modulated bottom and a linear confining potential. These potentials provide new tools to investigate the dynamics of 1D quantum systems and will allow us to address exciting questions in quantum thermodynamics and quantum simulations.
\end{abstract}

	\section{\label{sec:INTRO}Introduction}
	
Ensembles of ultra cold atoms  are an ideal and versatile tool to investigate a plethora of physical phenomena.
Their near-perfect isolation from the environment, paired with the availability of powerful techniques to probe and manipulate them, allows for the observation of coherent quantum evolution over long times with unprecedented detail \cite{bloch2008manybody,Langen2015annurev}.
When building such model systems, flexible and precise control over the confining potential often represents a key ingredient to access new phenomena. 
The realization of homogeneous systems, for example, can unveil effects otherwise masked by an inhomogeneous density distribution \cite{Gaunt2013UniformPot,Hueck2017,Mukherjee2017}, which allowed for the observation of recurrences in our isolated quantum many-body system \cite{Rauer2018}. Full local and dynamic control of the potential landscape can allow for the controlled excitation of collective modes \cite{Ville2018}, the realization of sonic black holes \cite{Lahav2010,Steinhauer2016} or exploring highly entangled topological states\cite{tai2017topological}.
This flexibility is often achieved through optical dipole potentials \cite{grimm1999dipoletrap} of shaped light fields realized by spatial light modulators (SLMs)~\cite{Gauthier2016DirectImaging, Ha2015RotonMaxon, Zupancic2016Holographic, Aidelsburger2017Relax2d}.

	\begin{figure}
		\centering
		\includegraphics[width=0.8\columnwidth]{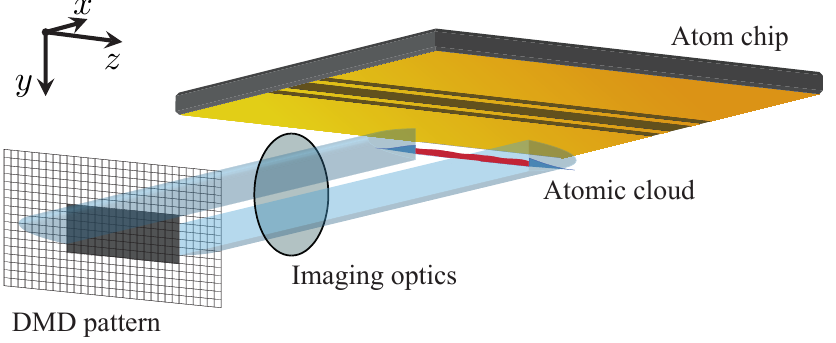}
		\caption{Combining optical dipole potentials with the magnetic confinement created by current carrying micro-wires of an atom chip. The DMD pattern is imaged onto the plane of the atomic cloud trapped below the atom chip. The controllable light field can be used to modify and `correct' the trapping potential created by the atom chip.
		In the remainder of the manuscript, the  $x$-axis is oriented along the propagation direction of the dipole light.
		The $y$-axis points towards gravity and the $z$-axis is oriented along the longitudinal direction of the atom chip trap. For simplicity, the DMD pattern shown in this schematic only cuts out the center of the elongated Gaussian beam and produces steep walls on the sides. For an example of an intensity profile correcting the potential roughness see Fig.~\ref{fig:detuning} and for a typical DMD pattern see Fig.~\ref{fig:dmd_pixel_change}.}
		\label{fig:atomchip_dmd}
	\end{figure}
	
Another versatile technique for trapping and manipulating cold neutral atoms are micro-traps generated by atom chips \cite{Folman2002,reichel2011traponchip}. In a number of experiments a large variety of potential configurations have been realized using combinations of static magnetic \cite{Reichel1999,folman2000firstatomchip}, electric \cite{Kruger2003}, optical \cite{Gallego2009} and radio-frequency (RF) fields~\cite{hofferberth2006rfdress}. 
Atom chips are well suited to create precisely controlled, steep and robust confinements, in particular highly elongated trapping configurations.
In such traps, the atoms are tightly confined along two spatial dimensions and loosely confined along the third dimension, realizing a quasi one-dimensional (1D) configuration (see Fig.~\ref{fig:atomchip_dmd}).
In our setup, the tight transverse confinement is controlled by the current flowing in a single micro-wire and a homogeneous external bias field. The weak longitudinal confinement, on the other hand, is generated by additional current carrying wires on the chip.
The detailed shape of the longitudinal confinement is influenced by imperfections in the current flow  \cite{krueger2007potentialroughness} which can be measured by magnetic field microscopy \cite{Wildermuth2005,Wildermuth2006,Aigner2008,Yang2016SQCRAMscope}.

In this paper we demonstrate that the spatial structure of the longitudinal trapping potential and its imperfections can be `corrected' by applying an additional structured dipole potential to the existing atom chip trap, which maintains the transverse trapping and manipulation. A far blue detuned light field will create a conservative repulsive potential for the atoms. Applying the structured light field from a direction orthogonal to the 1D chip trap allows nearly arbitrary modification of the potential landscape along the 1D trap.

This setup preserves the advantages of the tight magnetic trap in the transverse direction, such as the ability to deform the trap into a double-well potential using RF dressed state potentials, while greatly improving the control over the longitudinal 1D confinement.
We use a digital micro-mirror device (DMD) to shape arbitrary intensity profiles necessary for realizing a large variety of 1D optical dipole potentials.

	\section{\label{sec:IDEA} Basic design}

\begin{figure}
		\centering
		\includegraphics[width=0.9\columnwidth]{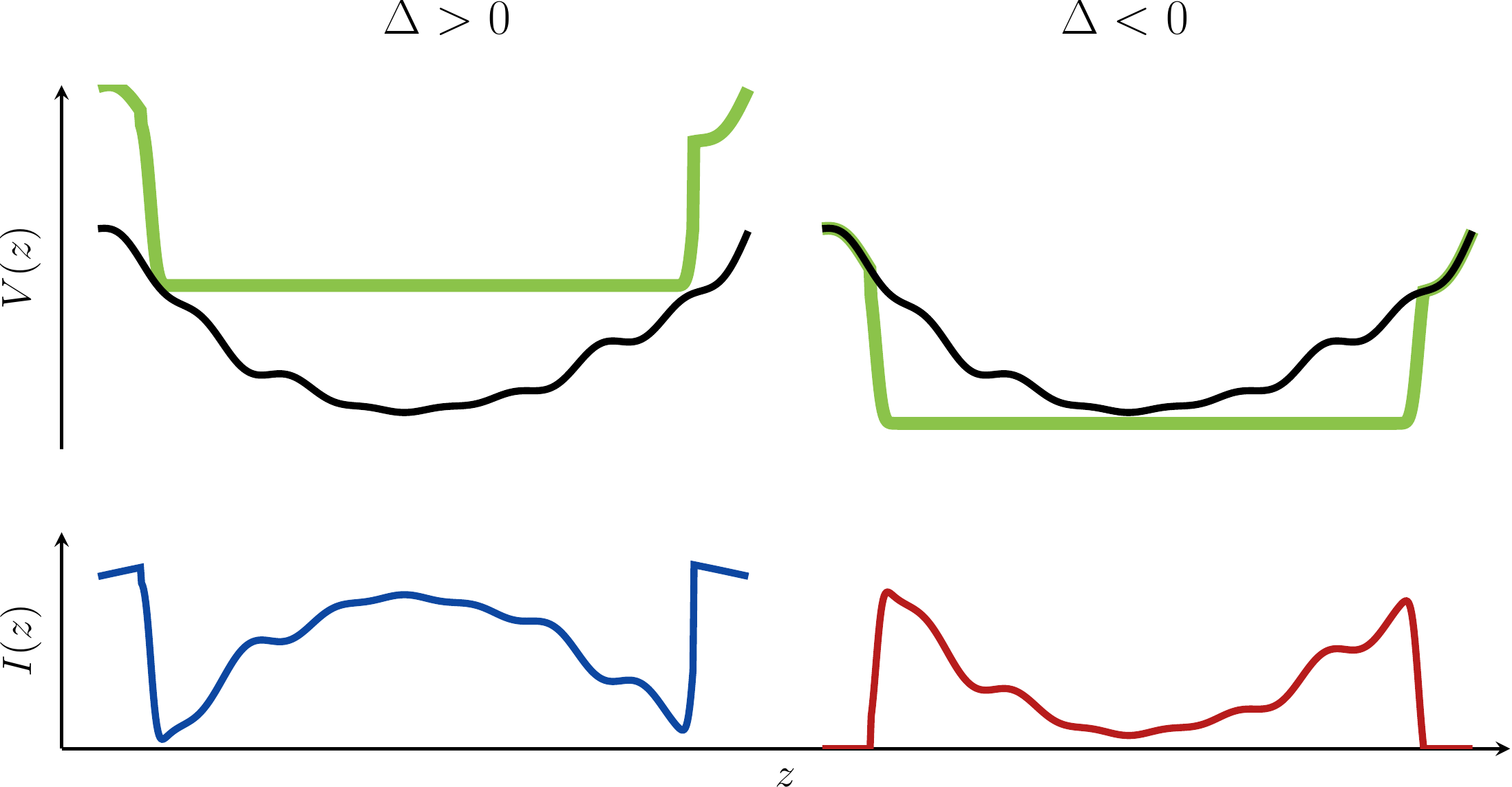}
		\caption{Creating a box-like potential (green) by superposing blue- or red-detuned light fields ($\Delta > 0$ and $\Delta < 0$ respectively) with a rough harmonic potential (black). The intensity profiles needed to achieve the target potential is shown for each case.}
		\label{fig:detuning}
	\end{figure}

	The potential created by a light field interacting with induced dipole moments of neutral atoms is proportional to the intensity of the light field and the sign of the detuning, $V_{\mathrm{dip}}(\bm{r}) \propto \mathrm{sgn}(\omega-\omega_0)I(\bm{r})$ ~\cite{grimm1999dipoletrap}. Here $\omega_0$ is the optical transition frequency, $\omega$ and $I(\bm{r})$ are driving frequency and intensity of the light field, respectively. 
	The sign of the detuning $\Delta = \omega-\omega_0$ determines whether the potential is repulsive ($\Delta > 0$, blue-detuned) or attractive ($\Delta < 0$, red-detuned). Figure~\ref{fig:detuning} shows an example of how the right intensity distribution can be used to create a box-like potential by correcting a rough harmonic confinement through the addition of blue- or red-detuned dipole potentials.
	
	The spatial structure of optical dipole potentials is determined by the spatial structure of the light beam. As such, beam-shaping techniques based on different types of SLMs can be exploited to modulate not only the intensity but also the phase of light fields. We use a DMD which is a reflective SLM offering fast and flexible intensity modulation of light fields.
	
	DMDs are powerful devices in terms of controllability and speed of performance, but ultimately, they can only produce pixelated binary patterns. To create gray-scale images from binary DMD patterns, time and/or spatial averaging has to be applied.
	
	Imaging the 2D binary pattern of a DMD results in the 2D intensity distribution $I(y,z)$ on the image plane (plane of atoms). Consider a 1D Bose gas extending along the $z$-axis in a highly anisotropic harmonic trap ($\omega_\perp / \omega_z \gg 1$). For typical experimental values of $\omega_\perp$, the transverse width of such an atomic cloud is much smaller than the resolution of commonly used optical imaging systems. For example, in our experiment with $\omega_\perp \sim 2\pi \times 2\, \si{\kilo \hertz}$, the width of the Gaussian ground state wave function is $a_\perp = \sqrt{\hbar/(M\omega_\perp)} \sim 250\, \si{\nano \meter}$, with $M$ being the mass of a \ce{^{87}_{}Rb} atom.
	Thus, assuming that the cloud is located at $y_\mathrm{cloud}$, only a tiny slice from the image, $I(z)\equiv I(y = y_\mathrm{cloud},z)$, is relevant for shaping the longitudinal potential.
    In the rest of the manuscript, $I(z)$ refers to a 1D intensity profile in the longitudinal direction at $(y = y_\mathrm{cloud},z)$.
    
\begin{figure}
		\centering
		\includegraphics[width=0.8\columnwidth]{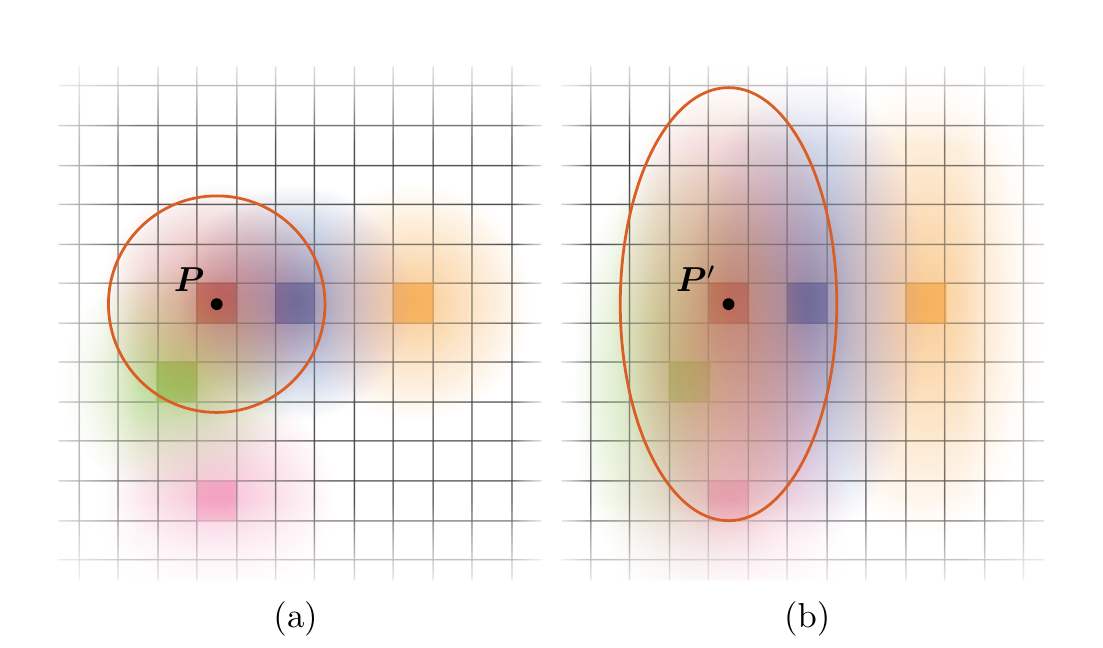}
		\caption[Expanding the dynamical range of intensity using the concept of spatial averaging]{Expanding the dynamical range of intensity using the concept of spatial averaging and spatial filtering. The images of single DMD pixels are shown in the image plane. For a diffraction limited spot much smaller than the pixel size,
		the DMD pixels are well resolved in the image plane, and of the five pixels switched on (shown in different colors), only one (red) would contribute to the intensity of point $\mathrm{P}$ ($\mathrm{P}^\prime$). If the image is blurred due to the finite imaging resolution,
		the pixels are imaged as broad, overlapping spots. In this case all pixels within the orange circle (ellipse) contribute to the intensity of point P ($\mathrm{P}^\prime$). In (a), the point spread function (PSF) of the imaging system is symmetric and three of the five active pixels (red, blue, green) contribute substantially to the intensity at point P. In (b), horizontal-pass spatial filtering broadens the PSF in the vertical direction, which leads to an increased number of pixels contributing to the intensity at point $\mathrm{P}^\prime$.}
		\label{fig:spatial_ave}
	\end{figure}

    To achieve a fine intensity resolution in the projected light pattern we must implement spatial averaging of the binary DMD pattern. 
    This can be accomplished by designing the imaging system in such a way, that through large demagnification, many binary pixels contribute to a single diffraction limited spot in the image plane (see Fig.~\ref{fig:spatial_ave}).
    In our case of controlling the longitudinal potential of a 1D system, the demagnification has an additional advantage: the DMD pixels contribute to the intensity at $y_\mathrm{cloud}$ depending on their vertical distance from the central axis, and therefore we obtain an improved intensity resolution at $y=y_\mathrm{cloud}$.
    
    An additional knob of control is spatial filtering: manipulating the Fourier components of light affects the optical resolution of the system. Blocking high frequency components worsens the optical resolution, which ultimately extends the contribution range of a single pixel in the binary pattern of the DMD. 
   
    In the case of 1D systems, only the spatial resolution along the longitudinal direction is relevant. Reducing the numerical aperture along the perpendicular direction will leave the resolution in the longitudinal direction unchanged. But more pixels from the perpendicular direction will contribute to the 1D intensity. This improves the tuneability of the potential at $y = y_\mathrm{cloud}$. The price to pay for performing spatial filtering is a loss in overall intensity.
    
    Beyond spatial averaging fast intensity modulation by the DMD can be exploited to extend the dynamical range of the created potential. This modulation has to be fast compared to the dynamical timescales of the trapped atoms, such that only a time-averaged potential is seen by the atoms~\cite{landau1976mechanics, bell2016timeaveragering}.
    
    An alternative to imaging the DMD directly is to employ it in the Fourier plane of an imaging system as a binary filter to modulate both phase and amplitude of the light field~\cite{zupancic2016holography}. However, if phase modulation is not necessary, direct imaging turns out to be more straightforward.
    
	\begin{figure*}
		\centering
		\includegraphics[width=\columnwidth]{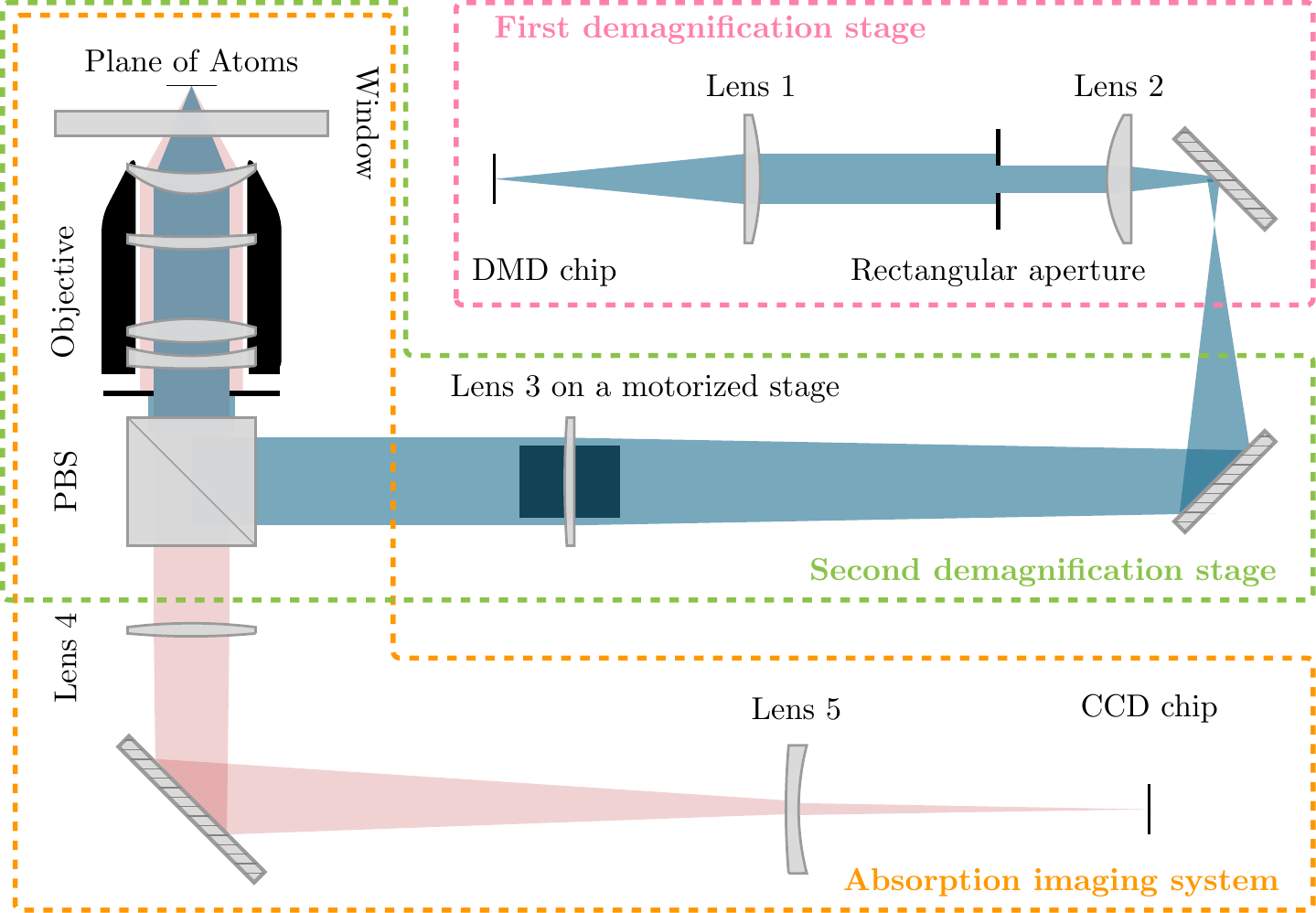}
		\caption[Integration of the optical setup for DMD shaped dipole potentials into the absorption imaging system (AIS)]
		{Integration of the optical setup for DMD shaped dipole potentials into the absorption imaging system (AIS).
		In the first demagnification stage, lens 1 (Thorlabs LA1256-B) along with lens 2 (Thorlabs LA1050-B) form a 4f system which is connected to the second demagnification stage by two mirrors. The first mirror has its backside polished and an overview camera placed behind it to observe the DMD pattern (not shown). Lens 3 (Thorlabs LA1727-B) together with the objective of the AIS forms the second demagnification stage. This lens can be shifted with a motorized stage to adjust the focus of the DMD imaging system in the plane of atoms. The dipole trap path is then superimposed to the imaging path on a polarizing beamsplitter (PBS) cube. While the p-polarized imaging light, $\lambda=780\, \si{\nano \meter}$ (pink) passes through the PBS cube, the s-polarized optical trap light, $\lambda=660\, \si{\nano \meter}$ (blue) is reflected into the home-built diffraction limited objective of the AIS. The objective consists of four commercially available lenses (Thorlabs LE1527, LE1359, AC508-150-B and LF1129) and has a NA of $0.26$~\cite{gring2012prethermalization}.
		The imaging beam is focused on the CCD chip via lens 4 (Thorlabs AC508-400-B) and lens 5 (ThorlabsLF1764) that form a telephoto group to shrink the size of the optical system.}
		\label{fig:setup_demags}
	\end{figure*}

	\section{\label{sec:IMPLEMENTATION}Experimental implementation}
	
	In our experiment, we realize 1D degenerate Bose gases of \ce{^{87}_{}Rb} atoms with an atom chip~\cite{folman2000firstatomchip}. Current carrying micro-fabricated wires on the chip surface along with an external bias field create a highly anisotropic Ioffe-Pritchard type trap with approximately 3D harmonic confinement: $V_{\mathrm{mag}}(\bm{r})= M\omega_\perp^2\,(x^2+y^2)/2 + M\omega_z^2\,z^2/2 $, with $M$ being the mass of a single atom, $\omega_\perp$ the transverse frequency and $\omega_z$ the longitudinal frequency of the trap. Typical parameters are $\omega_z \sim 2\pi \times 10\,\si{\hertz}$ and $\omega_\perp \sim 2\pi \times 2\,\si{\kilo \hertz}$. Furthermore, the cigar shaped cloud can be transversally split into two by applying RF fields~\cite{hofferberth2006rfdress}.

	Absorption imaging is used to obtain projected 2D atomic density distributions, $\rho(y_k, z_k)$, in time of flight (TOF) after switching off the trap~\cite{Smith2011imagingatomchip}. The spatial grid $(y_k, z_k)$ correspond to camera pixels in the plane of atoms.
	Density profiles along the longitudinal axis of the 1D system, $\rho(z_k)$, are extracted by integrating over the $y$-direction. The scheme of our absorption imaging system (AIS) is shown in Fig.~\ref{fig:setup_demags}. The projected size of the camera pixel in the object plane of the AIS (plane of atoms) is $1.05\, \si{\micro \meter}$. The influence of the imaging resolution of the AIS onto the measured 1D profiles can be approximately modeled by a convolution
	with a Gaussian point spread function with $\sigma_\mathrm{AIS} = 2.1\, \si{\micro \meter}$~\cite{schweigler2019thesis}.
	
	The small transverse width of the cloud renders direct \textit{in situ} imaging not practical. The weak longitudinal confinement suppresses the longitudinal expansion and allows us to measure the longitudinal density profile at short TOFs (between $1$ and $3\,\si{\milli \second}$) with negligible influence of the expansion. For cold samples, \textit{in situ} thermal fluctuations are also negligible at this short TOFs~\cite{imambekov2009drtheory, manz2010dr,Nieuwkerk2018ProjectivePhase1d}.
	
	Although a variety of trapping geometries can be produced using RF dressed state potentials, once the atom chip is designed, adjusting the longitudinal confinement is greatly limited. Moreover, there exists potential roughness due to imperfections of micro-wires on the chip~\cite{krueger2007potentialroughness}. 
	Compensating for this potential roughness without an additional potential is possible in principle, but requires special chip designs~\cite{trebbia2007roughnessSuppress}.
	Adding an optical dipole potential allows to correct the potential roughness and shape the longitudinal trapping potential in a very flexible way.  
	
	To create the dipole potentials we use 
	a $\lambda_\mathrm{dip} = 660\, \si{\nano \meter}$ laser (opus660, Laser Quantum) which provides $1.5\, \si{\watt}$ of spatially single mode light, blue-detuned with respect to the D line of \ce{^{87}_{}Rb} ($\lambda_\mathrm{D_2} = 780\, \si{\nano \meter}$, $\lambda_\mathrm{D_1} = 795\, \si{\nano \meter}$).
	For our purpose, which is primarily correcting an existing potential and creating potentials with steep walls in 1D, a blue-detuned light field is advantageous~\cite{grimm1999dipoletrap}.
	The choice of $\lambda_\mathrm{dip}$ is mainly motivated by the anti-reflection coating of the existing optical elements in the experiment (see Fig.~\ref{fig:setup_demags}) which favours wavelengths ranging from $650$ to $1150\,\si{\nano \meter}$. Being far-detuned from the resonance guarantees that for $U_{\mathrm{dip}}/ \hbar \sim 2\pi \times 1\, \si{\kilo \hertz}$ the scattering rate is $\sim 2\pi \times 0.1\, \si{\milli \hertz}$. This leads to only one spontaneous scattering per second for an ensemble of $10^4$ atoms, causing negligible atom loss and heating.
	
	We apply the concepts discussed in Sec.~\ref{sec:IDEA} by imaging DMD patterns directly onto the plane of atoms. Our DMD chip is a DLP9500 from Texas Instruments with $1080\times1920$ micro-mirrors, each $10.8\times 10.8\,\si{\micro \meter}$. The refresh rate for the whole array is $17\,\si{\kilo \hertz}$, which increases to $50\,\si{\kilo \hertz}$ when the number of active rows is limited to $50$. The DMD chip is a part of the V-9501 module by ViALUX. This module also contains a V4395 main board which includes a completely configured high-speed FPGA logic and firmware. 
	
	The optical setup imaging DMD patterns onto the plane of atoms is shown in Figure \ref{fig:setup_demags}. It consists of two demagnification stages: first, a 4f system is used to perform the horizontal-pass spatial filtering using an adjustable rectangular slit. Secondly, a plano-convex lens along with the objective of the AIS images the DMD pattern onto the plane of atoms.
	
	The DMD chip is placed at the focal point of the first lens and the distance between lens 1 and lens 2 is the sum of the two focal lengths, $f_1 + f_2 = 30\, \si{\centi \meter} + 10\, \si{\centi \meter} = 40\, \si{\centi \meter}$. The width of the rectangular slit sitting in the Fourier plane of the system can be adjusted independently in vertical and horizontal direction from $0$ to $12\,\si{\milli \meter}$.
	The image formed by the first two lenses is further demagnified by lens 3 and the objective of the AIS. The focus of the whole system is adjusted by lens 3 which is mounted on a motorized stage. 
	This is necessary such that the AIS and the dipole light can be focused independently.
	The total magnification of the DMD imaging system is $1/25.9$ and it has a minimum spot size (Airy diameter) of $4\,\si{\micro \meter}$ in the plane of atoms.  This implies that more than 70 DMD pixels contribute to the Airy disk at a single point.
	The DMD chip is illuminated by a Gaussian beam whose horizontal extension is elongated by a cylindrical telescope.
	This is done to match the available intensity to the size of the cloud and thereby efficiently use the available power. 
	The polarization of the dipole potential beam is also set such that the beam gets reflected in the PBS cube in the second demagnification stage with minimal loss.

	\section{\label{sec:OPTIMIZATION}Real-time pattern optimization}

	With the optical setup described in the previous section, an area of interest (AOI) on the DMD chip with $n_y\times n_z$ pixels (typically $20\times 500$ pixels for a $200\, \si{\micro \meter}$ long 1D trap) will be imaged onto the plane of atoms. With these pixels, $2^{(n_y\times \, n_z)}$ different patterns can be created. The challenge is to find a pattern which produces a 1D optical dipole potential which results in the desired target potential $V_\mathrm{T}(z)$ when superposed with the magnetic confinement of the atom chip. In principle, we can calculate the 2D image of DMD pattern numerically and design the pattern to obtain desired potentials. However, due to interferences the actual dipole potential is sensitive to precise alignment of optical elements and imperfections in the setup and is difficult to predict precisely in a real world setting. Hence, an iterative pattern optimization is used to obtain the best possible reconstruction of the target potential.
	
	In the experiment, we measure density profiles $\rho(z_k)$. Thus, the ultimate goal of our pattern optimization process is to find a DMD pattern which creates a 1D density profile $\rho(z_k)$ as close as possible to a simulated target density $\rho_\mathrm{T}(z_k)$. The target density is obtained by calculating the ground state of a 1D non-polynomial Schr{\"o}dinger equation (NPSE) \cite{salasnich2002effective1d} with the external target potential $V_\mathrm{T}(z_k)$.
	
\begin{figure}
		\centering
		\includegraphics[width=\columnwidth]{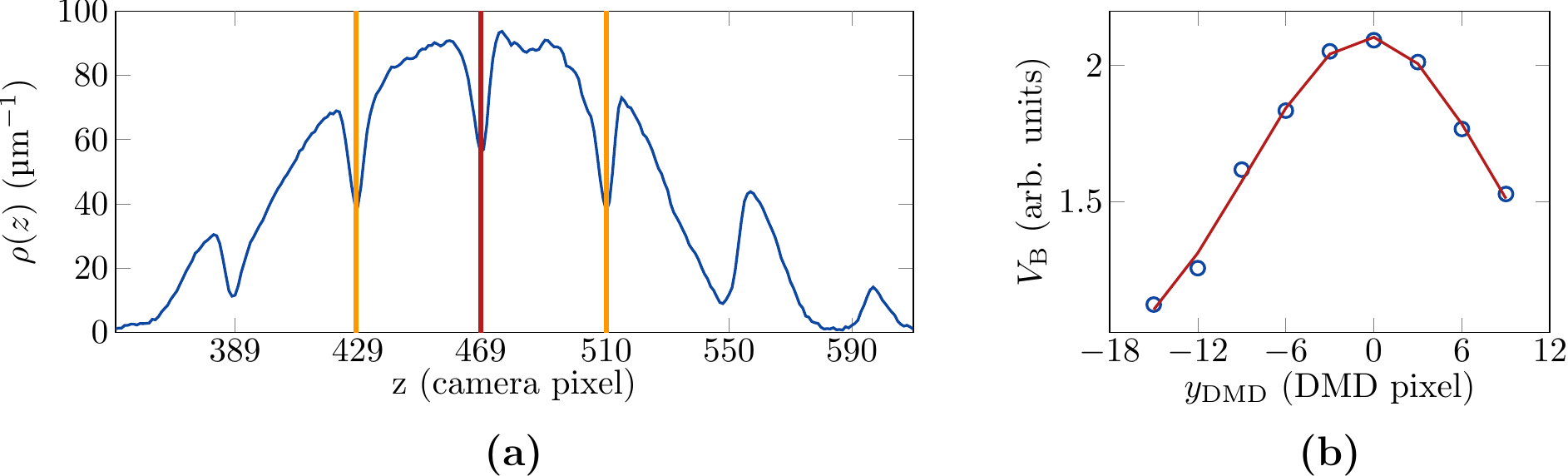}
    		\caption{An example of data used for (a) horizontal (longitudinal) calibration and (b) vertical calibration. (a) multiple super pixels ($190\times \, 3$ pixels) are turned on $100$ DMD pixels apart from each other. The red vertical line corresponds to the image of the super pixel in the center of the DMD pattern. The distance between two orange lines and the position of the red line can be used for a linear mapping between DMD pixels and corresponding camera pixels. (b) height of the barrier ($V_\mathrm{B}$) created by a super pixel ($3\times \, 15$ pixels) is plotted for different vertical positions of the super pixel on the DMD pattern. The row $y_\mathrm{DMD} = 0$ is the central DMD row. The blue circles represent the data taken from $y_\mathrm{DMD} = -15$ to $9$ with steps of $3$ DMD pixels and the red curve is a Gaussian fit.
		}
		\label{fig:calibration}
	\end{figure}

	Before starting the pattern optimization, a calibration is performed to infer the corresponding position in the plane of atoms for the different DMD pixels in horizontal and vertical directions separately. For horizontal (longitudinal) calibration, a number of DMD columns are turned on at different positions. Their position is then detected on the 1D density and used to derive a linear map (Fig.~\ref{fig:calibration}(a)). To perform vertical calibration, a simple barrier is created by a thin super pixel in the center of the atomic cloud (typically $3$ rows with $3$ to $50$ pixels each). Shifting this super pixel vertically changes the barrier height. At the point where this height is maximum, the center of the super pixel is selected as the central row of the DMD pattern (Fig.~\ref{fig:calibration}(b)). This row corresponds to the position of the elongated atomic cloud ($y=y_\mathrm{cloud}$) in the plane of atoms and has the maximum contribution to $I(z)$. The vertical calibration also defines the vertical AOI, the area outside where the DMD pixels do not contribute to $I(z)$. 
	
	The process of optimizing the DMD pattern starts by choosing an initial pattern $p^0$. This pattern can be either totally black (all DMD pixels are off), a pattern that was calculated to correct the actual potential $V(z_k)$ estimated from a measurement of $\rho(z_k)$, or a pattern from a previous optimization processes.
	In each iteration $i$, with a DMD pattern $p^i$, the average 1D density profile, ${\bar{\rho}}^i(z_k)$, and the mean atom number, $\bar{N}^i$, are obtained from $n$ absorption pictures. The mean atom number $\bar{N}^i$ is then used to calculate $\rho_\mathrm{T}^i(z_k)$.
	
	To estimate how well the measured density ${\bar{\rho}}^i(z_k)$ and the calculated target density $\rho_\mathrm{T}^i(z_k)$ agree, we calculate the normalized root-mean-square deviation,
    \begin{equation}\label{eq:e_rms}
    	\epsilon_\mathrm{RMS}^i = \sqrt{ \frac{1}{j-l} \sum_{k=j}^l \left( \frac{\rho_\mathrm{T}^i(z_k) - {\bar{\rho}}^i(z_k)}{\rho_\mathrm{T}^i(z_k)}\right)^2 } \, ,
    \end{equation}
    in each iteration for a specific region from $z_j$ to $z_l$ and compare it with a target value, $\epsilon_\mathrm{T}$.
    Ultimately, the quality required for a specific target potential, depends on the application.
    Note that however, there is a lower limit for $\epsilon_\mathrm{T}$ which is given by the noise in the density measurement: $\epsilon_\mathrm{T}$ should not be smaller than the average of the standard error of the mean density, $\sigma_{\bar{\rho}}$, in the same region.
	
We terminate the optimization process when $\epsilon_\mathrm{RMS}^i < \epsilon_\mathrm{T}$. In that case, the final DMD pattern is $p^i$. If $\epsilon_\mathrm{RMS}^i \geq \epsilon_\mathrm{T}$ a new pattern $p^{i+1}$ is generated using an update procedure discussed below and then the process is repeated. 
	
	In the DMD pattern update procedure, the states of individual pixels are altered based on the difference between $\bar{\rho}^i(z_k)$ and $\rho_T^i(z_k)$. In regions where the measured density is higher than the target density more pixels are switched on, thus yielding a locally higher potential. Conversely, by switching pixels off or shifting active pixels away from the center, the optical dipole potential in the corresponding region is decreased resulting in higher densities.
	The logic behind the optimization is based on the one-to-one correspondence between the trapping potential and the (Thomas-Fermi) density distribution (see Eq.~(\ref{eq:V_TF})). In a trapped degenerate ensemble with constant number of atoms, locally increasing (decreasing) the potential decreases (increases) the local density distribution.

	\begin{figure*}
		\centering
		\includegraphics[width=0.95\columnwidth]{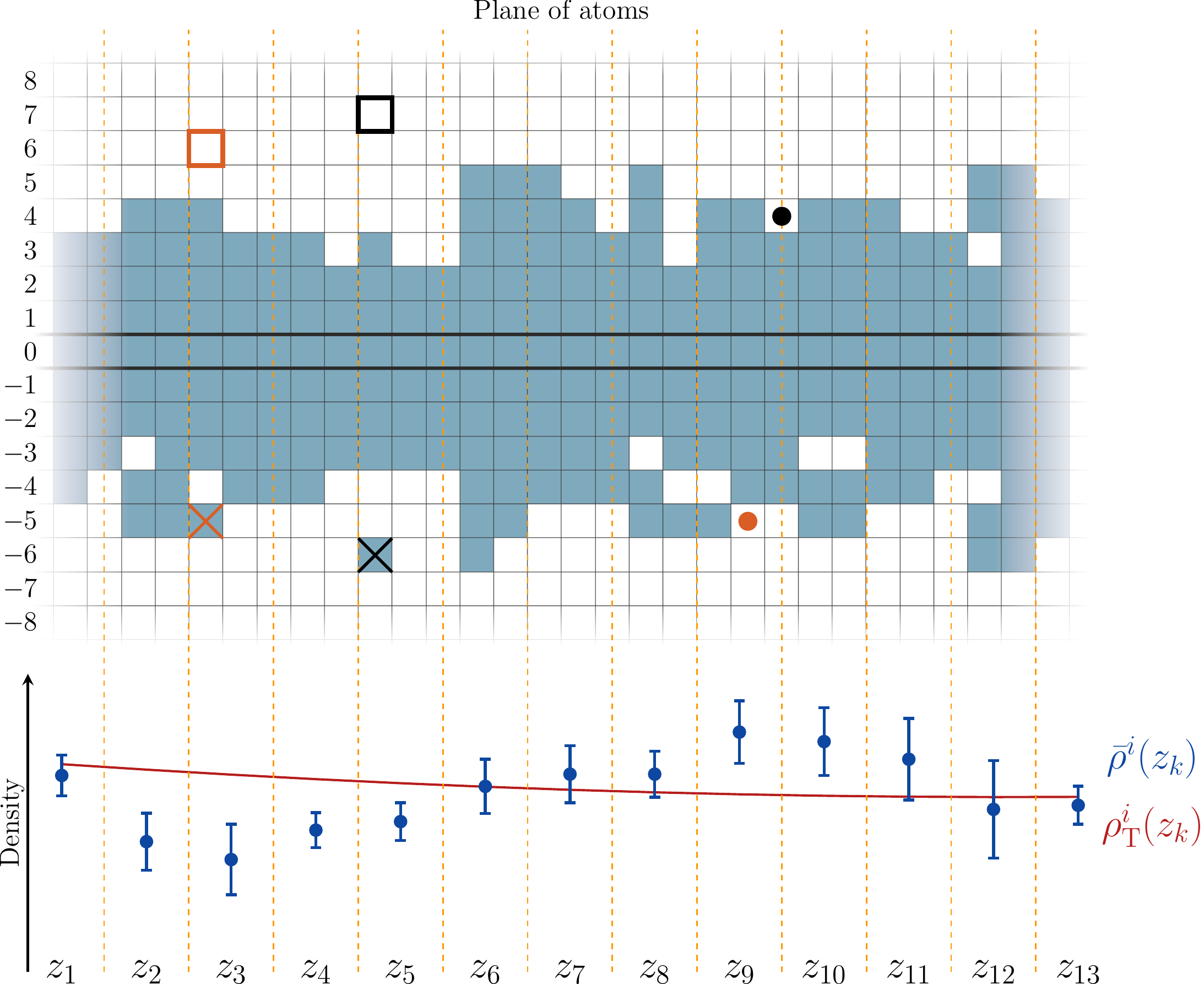}
		\caption{Updating the DMD pattern based on the comparison between the averaged measured density, $\bar{\rho}^i(z_k)$, and the target density, $\rho_T^i(z_k)$. The process is explained in detail in the final part of Sec.~\ref{sec:OPTIMIZATION}. The example here shows a small longitudinal segment of iteration number $i =14$ of the optimization leading to the results shown in Fig.~\ref{fig:all_den_v}(c). The region on the DMD corresponds to an interval $z_1= -39.89\, \si{\micro \meter}$ to $z_{13}= -26.23\, \si{\micro \meter}$.}
		\label{fig:dmd_pixel_change}
	\end{figure*}
	 
	In Fig.~\ref{fig:dmd_pixel_change}, the image of seventeen rows of the DMD chip is shown in the plane of atoms, where the zeroth row is the central row discussed before. The orange dashed lines represent the camera pixels mapped in the plane of atoms, defining the spatial grid points $z_k$. The length of each grid point ($1.05\, \si{\mu \meter}$) is equivalent to the length of almost $2.5$ DMD pixels mapped in the plane of atoms ($2.5\times 10.8\, \si{\mu \meter} / 25.9 = 1.04\, \si{\mu \meter}$).
	
	For each grid point $z_k$, the corresponding region on the DMD remains unchanged if $|\bar{\rho}^i(z_k) - \rho_T^i(z_k)| < \sigma^i_{\bar{\rho}}(z_k)$ where $\sigma^i_{\bar{\rho}}(z_k)$ is the standard error of the mean density, represented by the error bars in this figure. In our example here, this is the case for $z_1$, $z_6$, $z_7$, $z_8$, $z_{11}$, $z_{12}$ and $z_{13}$. In the other cases where $|\bar{\rho}^i(z_k) - \rho_T^i(z_k)| > \sigma^i_{\bar{\rho}}(z_k)$, the magnitude of the total deviation determines how many pixels corresponding to grid point $z_k$ will be changed.
	\begin{figure*}
		\centering
		\includegraphics[width=\columnwidth]{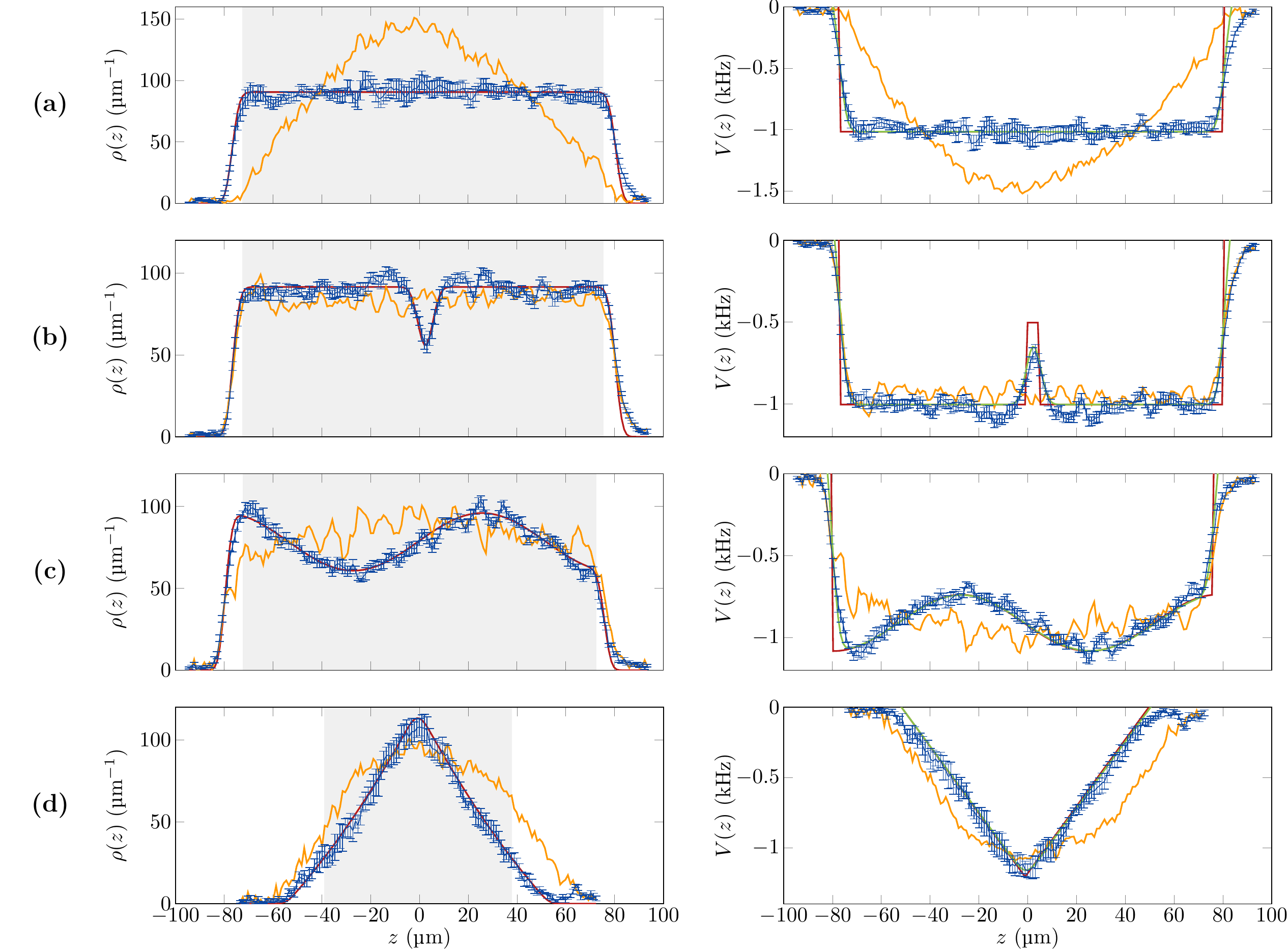}
		\caption{Results of pattern optimization process for four different cases: (a) a box-like potential with length $L = 160\, \si{\micro \meter}$, (b) two box-like potentials separated by a barrier, (c) a $L = 160\, \si{\micro \meter}$ box-like potential with a sinusoidally modulated bottom and (d) a V-shaped potential. In the left (right) column, orange curves are initial averaged 1D densities (potentials), red curves are target densities (potentials) and blue curves are averaged 1D density profiles (potentials) and their standard errors for optimized patterns. The green curves in the right column are the target potentials broadened by the DMD imaging system. The measured potentials show a good agreement with these broadened target potentials. Note that the scale of density (potential) axis is not the same for different cases. For the final density profiles, $\epsilon_\mathrm{RMS}$ calculated over the gray shaded region is (a) $4.4\%$, (b) $5.5\%$, (c) $4.2\%$ and (d) $5.9\%$. For (b) the optimization process started with the optimized pattern achieved in (a). For (c) the optimization started with an older optimized pattern for a box-like potential which had a slightly different calibration settings. In (a) and (d) however, the optimization process started with a pattern in which all DMD pixels were off, i.e. orange curves in (a) and (d) represent our longitudinal magnetic potential when the condensate is split transversally in a double-well. Since the measured density for the first iteration is only averaged over two experimental realizations, the effect of shot noise is evident in the orange curves.
		}
		\label{fig:all_den_v}
	\end{figure*}
	
	If $\bar{\rho}^i(z_k) - \sigma^i_{\bar{\rho}}(z_k) > \rho_T^i(z_k)$, the potential has to be increased by turning additional pixels on. To chose these, first the distances of the outermost active pixels for the columns involved are compared together to select the column for which this distance is the smallest. Consider three columns involved in the region defined by $z_{10}$: the maximum distance of an active pixel from the central row is only 4 for the column in the left. However, for the central column and for the column in the right this number is $5$. So in this case the left column will be chosen. In the selected column, if the outermost pixel is located below the central row, the pixel above the central row which is one row further away will be turned on, if the outermost pixel is above the central row, then the pixel with the same distance but below the central row will be turned on. For $z_{10}$, the pixel marked with the black dot will be turned on. For $z_9$, the pixel marked by the orange dot will be turned on.
	
	If $\bar{\rho}^i(z_k) + \sigma^i_{\bar{\rho}}(z_k) < \rho_T^i(z_k)$, the potential has to be decreased by shifting the outermost pixel away from the central row in each region. For example in the region corresponding to $z_3$ ($z_5$) the pixel in the row number $-5$ ($-6$) will be shifted to the row number $6$ ($7$) in the same column, as marked in the Fig.~\ref{fig:dmd_pixel_change}. Note that in principle, the pixel $-5$ ($-6$) could also be shifted to the row $-6$ ($-7$). Although due to the interference results may be different, choosing one over the other is arbitrary. If the selected pixel is in the border of AOI it will be simply turned off.

	\begin{figure*}
		\centering
		\includegraphics[width=\columnwidth]{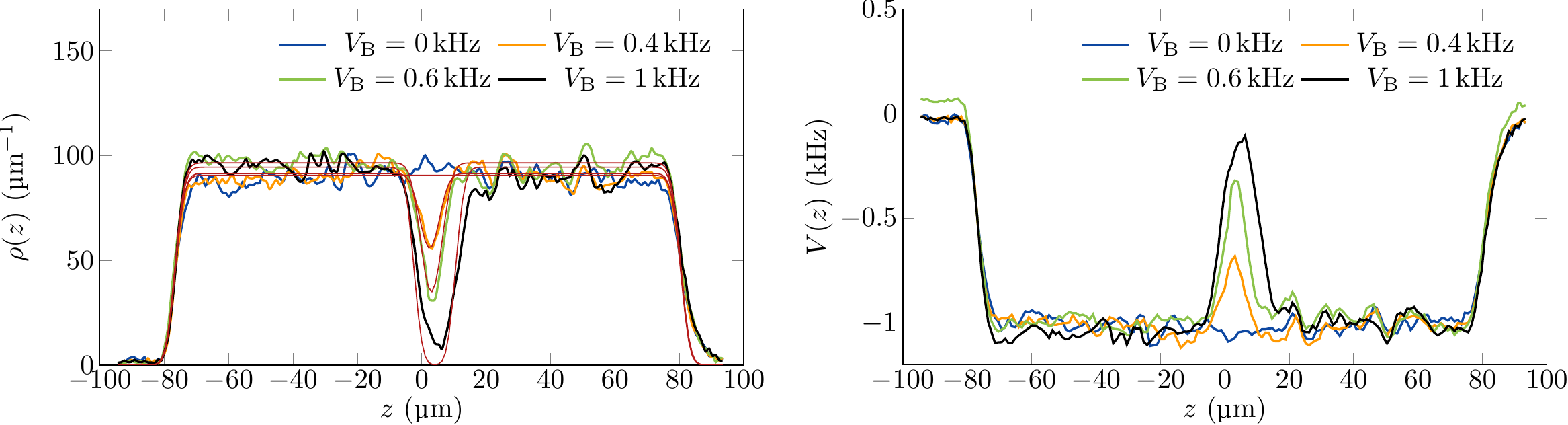}
		\caption{Realization of a $L = 160\, \si{\micro \meter}$ long box-like potential with different barrier heights, $V_\mathrm{B}$, from $0$ to $1\, \si{\kilo \hertz}$. Density profiles and potentials are shown with different colors. Thin red curves represent the target densities for each case. The figure suggests that the widths of barriers of the measured densities agree with the simulated target densities.
		}
		\label{fig:barriers}
	\end{figure*}

	\begin{figure}
		\centering
		\includegraphics[width=0.7\columnwidth]{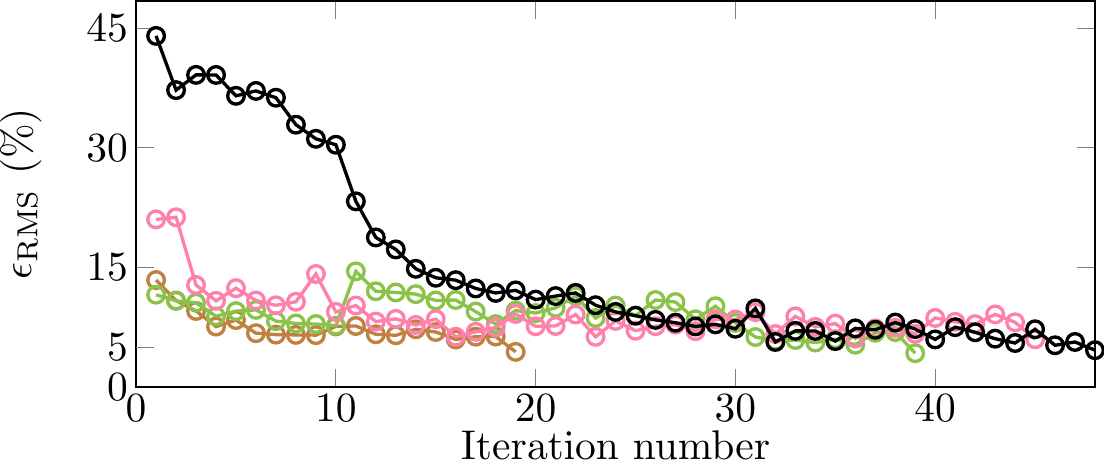}
		\caption{Evolution of $\epsilon_\mathrm{RMS}$ for three different cases (lines are guides to the eye): a box-like potential with length $L = 160\, \si{\micro \meter}$ (black), two box-like potentials separated by a barrier (brown), a box-like potential with a sinusoidally modulated bottom (green) and a V-shaped potential (pink). All initial and final densities and potentials are plotted in Figs.~\ref{fig:all_den_v}(a)-\ref{fig:all_den_v}(d).
		}
		\label{fig:errors}
	\end{figure}
	
	\section{\label{sec:EXAMPLES}Results}
	We used the technique described in Sec.~\ref{sec:OPTIMIZATION} to realize a variety of 1D trapping potentials including a box-like potential, two box-like potentials separated with barriers with different heights, a box-like potential with sinusoidal modulations and a V-shaped potential. Results are shown in Figs.~\ref{fig:all_den_v} and \ref{fig:barriers}.
	
	In these examples, the optimization process started with different initial DMD patterns. In the first and the fourth example (Figs.~\ref{fig:all_den_v}(a) and \ref{fig:all_den_v}(d)), the optimization started with all DMD pixels turned off. Thus the initial density profiles (potential) plotted in Figs.~\ref{fig:all_den_v}(a) and \ref{fig:all_den_v}(d) (orange) represent the 1D magnetic potential. In the examples presented in Figs.~\ref{fig:all_den_v}(b) and \ref{fig:all_den_v}(c)  we  started with a previously optimized DMD pattern.
	
The initial and final potentials, plotted in Fig.~\ref{fig:all_den_v}, are obtained using the NPSE for 1D condensates ~\cite{salasnich2002effective1d}. In the limit where the kinetic energy of motion in $z$-direction is neglected (Thomas-Fermi approximation), the solution for the stationary state reads:  
	\begin{equation}\label{eq:V_TF}
    	V(z)  = \mu -\frac{g}{2\pi a_\perp^2} \frac{\rho(z)}{\sqrt{1+2a_\mathrm{s} \rho(z)}} -\frac{\hbar \omega_\perp}{2} \left( \sqrt{1+2a_\mathrm{s} \rho(z)} + \frac{1}{\sqrt{1+2a_\mathrm{s} \rho(z)}} \right)  \, .
    \end{equation}
	Here, $\rho(z)$ is the measured 1D density,  $a_\mathrm{s}$ is the s-wave scattering length, $g=4\pi \hbar^2 a_\mathrm{s}/M $ is the 3D scattering amplitude, $a_\perp$ is the oscillator length mentioned in Sec.~\ref{sec:IDEA}, and $\mu$ is the chemical potential which constitutes only an offset. Although this measured potential is not used in the optimization process, a qualitative comparison between the measured potential and the target potential is instructive.
	
	For the examples presented here, the number of experimental realizations was varied in each iteration depending on the normalized root-mean-squared deviation of the previous step. This helps speed up the optimization process in the beginning where only a rough measurement of the density is needed. At the initial stages of the process, only $n=2$ shots were used to calculate the averaged 1D density and the mean atom number. At the more advanced stages of the optimization this number was increased to $n=5$ and $n=10$ accordingly. For all measurements, the optical dipole trap was turned on during the evaporative cooling in a transverse double-well. Absorption pictures are taken after $2\,\si{\milli \second}$ TOF. 
	
	In Fig.~\ref{fig:errors}, the normalized root-mean-squared deviation in each iteration, $\epsilon^i_\mathrm{RMS}$, is plotted for different examples.
	The number of iterations needed to fulfill the termination criterion was different depending on initial and target potentials. For example, a local modification to an existing box-like potential (adding a barrier) was done faster than achieving the box-like potential starting with a harmonic potential.
	Given the fact that each experimental realization takes approximately $25\,\si{\second}$, the total optimization time for these examples was between $60$ to $135$ minutes.
	
\begin{figure}
		\centering
		\includegraphics[width=\columnwidth]{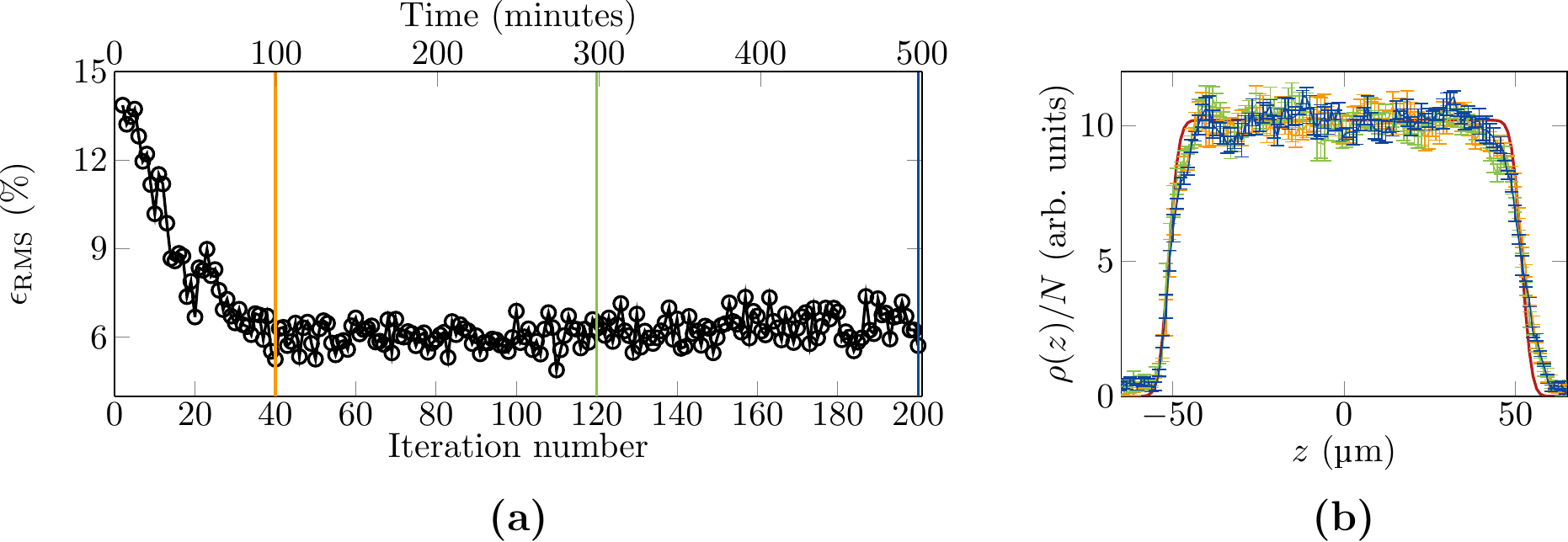}
		\caption{Stability measurement: (a) evolution of $\epsilon_\mathrm{RMS}$ before and after optimization for a box-like potential with length $L = 100\, \si{\micro \meter}$. The horizontal axis on the top is the time corresponding to different iterations. The optimization stopped at iteration number $40$ (orange line) and the DMD pattern is kept unchanged through out the measurement. (b) normalized target density (red), normalized density profiles for iteration numbers $40$ (orange), $120$ (green) and $200$ (blue) are plotted.
		}
		\label{fig:stability}
	\end{figure}

	To check the stability of achieved potentials after optimization, $\epsilon^i_\mathrm{RMS}$ is plotted over a time span of more than six hours in Fig.~\ref{fig:stability}(a). In this case, the optimization stopped after $40$ iterations and the achieved pattern remained unchanged through the rest of the measurement. In Fig.~\ref{fig:stability}(b), the orange curve is the normalized density measured immediately after the optimization stopped.
	Fig.~\ref{fig:stability}(b) shows a good agreement between normalized densities after $200$ (green), $400$ minutes (blue) and the orange curve.
	
	\section{\label{sec:CONCLUSION}Conclusion}
	
	We have implemented a robust and versatile way to control the longitudinal confinement of 1D Bose gases created on an atom chip. This is accomplished by modifying the longitudinal potential landscape defined by the magnetic fields by adding an optical dipole potential that is locally controlled by a DMD. We have furthermore demonstrated an autonomous optimization procedure to achieve any given longitudinal potential. 
	
	Our system opens up many new possibilities for experiments with 1D quantum systems. A box-like potential allows to create homogeneous systems, additional barriers will allow longitudinal splitting and tunneling. Switching between patterns will allow quench experiments. For example, imprinting a density modulation corresponding to the density quadrature of a specific mode will allow to excite this mode precisely and look at its evolution and decay. Furthermore, DMD patterns can be changed faster than the typical time scales of the 1D quantum gas. This will allow us to implement time varying potentials. One application will be to build and explore thermal quantum machines with 1D quantum fields~\cite{Schmiedmayer2018ThermalMachine}.

	\section*{\label{sec:Funding} Funding}
	
	Fonds zur F\"orderung der wissenschaftlichen Forschung (FWF) (I 3010-N27, 3980-N27, W 1210-N25);
	John Templeton foundation (JTF) (60478);
	Wiener Wissenschafts- und Technologie Fonds  (WWTF) (MA16-066);
	Funda\c{c}{\~a}o para a Ci\^{e}ncia e Tecnologia, Portugal (PD/BD/128641/2017); 
	\"{O}sterreichische Akademie der Wissenshaften (\"{O}AW) (ESQ Fellowship 801110)


	\section*{\label{sec:Ack} Acknowledgment }
	
	The authors would like to thank Kaspar Sakmann who contributed to the initial phase of this project.
	We also thank Andrew N. Kanagin for helping proof read the manuscript.

	\bibliography{Refs/references}

\end{document}